\documentclass{ws-ijmpcs}
\def\be{\begin{eqnarray}}
\def\ee{\end{eqnarray}}
\def\bea{\begin{eqnarray}}
\def\eea{\end{eqnarray}}

\def\0T{{\bf 0}_\perp}
\def\xbj{x_{\mbox{\tiny B}}}

\begin{document}

\markboth{Matthias Burkardt}{Quark Orbital Angular Momentum and Final State Interactions}

%
\catchline{}{}{}{}{}
%

\title{QUARK ORBITAL ANGULAR MOMENTUM AND FINAL STATE INTERACTIONS
}

\author{MATTHIAS BURKARDT
}

\address{
Department of Physics, New Mexico State University\\
Las Cruces, NM 88003, U.S.A.\\
burkardt@nmsu.edu}

\maketitle

\begin{history}
\received{Day Month Year}
\revised{Day Month Year}
\end{history}

\begin{abstract}
Definitions of orbital angular momentum based on Wigner distributions
are used to discuss the connection between the Ji definition
of the quark orbital angular momentum and that of Jaffe and Manohar. The difference between these two definitions can be interpreted as the change in the quark orbital angular momentum as it leaves the target in a DIS experiment.
The mechanism responsible for that change is similar to the mechanism that
causes transverse single-spin asymmetries in semi-inclusive deep-inelastic scattering.

\end{abstract}

\ccode{PACS numbers: 11.25.Hf, 123.1K}
\section{Introduction}
Generalized Parton Distributions (GPDs) have been identified as a powerful tool to analyze the
angular momentum decomposition of the nucleon\cite{JiPRL}.
Furthermore GPDs can also be used to create truly three-dimensional images of the nucleon in the form of impact parameter dependent parton distributions\cite{mbGPD}. These images in a space where one dimension describes the light-cone
momentum fraction and the other two dimensions describe the transverse position of the parton (relative to the
transverse center of momentum) are complemented by Transverse Momentum dependent
parton Distributions (TMDs)\cite{ams}. Wigner distributions provide a framework that allows a simultaneous description of GPDs and TMDs\cite{wigner}.

Orbital Angular Momentum (OAM) correlates the position and momentum of partons. One can thus utilize Wigner distributions, which
simultaneously embody the distribution of position and momentum,  to define OAM\cite{lorce,jifeng}.  
However, in the definition of these distributions,
care must be applied to ensure gauge invariance. This can be accomplished by connecting any nonlocal correlation function with a Wilson-line gauge link, which requires specifying a path along which the vector potential is evaluated.
The choice of path raises the immediate issue of how the quantities defined using
Wigner distributions (TMDs, OAM, ...) depend on that choice. The importance of this issue had become evident in the context of Single-Spin Asymmetries (SSAs) \cite{BHS}. Indeed, while a straight-line gauge link definition of TMDs yields a vanishing Sivers effect \cite{sivers,collins1}, the correct gauge link relevant for TMDs in
Semi-Inclusive Deep-Inelastic Scattering (SIDIS) involves a detour to light-cone infinity
\cite{jifengTMD} in order to properly include final-state interactions. In light-cone gauge, this subtlety had first been overlooked since in that gauge the Sivers effect solely arises from the contribution from the gauge-link piece at light-cone infinity \cite{jifengTMD}.

With Wigner distributions and OAM defined through them these issues arise all over again
\cite{jifeng,hatta,lorce2}. The main goal of this note is to address that dependence of OAM defined through Wigner distributions on the choice of path for the gauge link and 
to interpret the resulting difference between common definitions
of OAM.

\section{Transverse Single-Spin Asymmetries}

In a target that is polarized transversely ({\it e.g.} vertically), 
the quarks in the target 
can exhibit a (left/right) asymmetry of the distribution 
$f_{q/p^\uparrow}(\xbj,{\bf k}_T)$ in their transverse 
momentum ${\bf k}_T$ \cite{sivers,trento,RPP}
\be
f_{q/p^\uparrow}(\xbj,{\bf k}_T) = f_1^q(\xbj,k_T^2)
-f_{1T}^{\perp q}(\xbj,k_T^2) \frac{ ({\bf {\hat P}}
\times {\bf k}_T)\cdot {\bf S}}{M},
\label{eq:sivers}
\ee
where ${\bf S}$ is the spin of the target nucleon and
${\bf {\hat P}}$ is a unit vector opposite to the direction of the
virtual photon momentum. The fact that such a term
may be present in (\ref{eq:sivers}) is known as the Sivers effect
and the function $f_{1T}^{\perp q}(\xbj,k_T^2)$
is known as the Sivers function.
The latter vanishes in a naive parton 
picture since $({\bf {\hat P}} \times {\bf k}_T)\cdot {\bf S}$ 
is odd under naive time reversal (a property known as naive-T-odd), 
where one merely reverses
the direction of all momenta and spins without interchanging the
initial and final states. 
The significant distortion of parton distributions in impact 
parameter space \cite{IJMPA}
provides a natural mechanism for a Sivers effect.
In semi-inclusive DIS, when the 
virtual photon strikes a $u$ quark in a $\perp$ polarized proton,
the $u$ quark distribution is enhanced on the left side of the target
(for a proton with spin pointing up when viewed from the virtual 
photon perspective). 
\begin{figure}
\unitlength1.cm
\begin{picture}(10,2.3)(3.,19.2)
\includegraphics{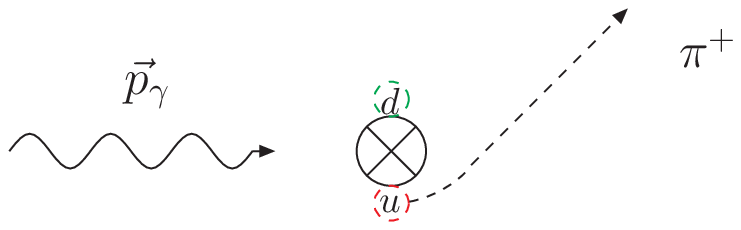}
\end{picture}
\caption{The transverse distortion of the parton cloud for a proton
that is polarized into the plane, in combination with attractive
FSI, gives rise to a Sivers effect for $u$ ($d$) quarks with a
$\perp$ momentum that is on the average up (down).}
\label{fig:deflect}
\end{figure}
Although in general the final state 
interaction (FSI) is very complicated, we expect it to be on average attractive thus translating a position space
distortion to the left into a momentum space asymmetry to the right
and vice versa (Fig. \ref{fig:deflect})\cite{mbSSA,Metz}. 
With this simple mechanism
one finds that $f_{1T}^{\perp u}<0$, while 
$f_{1T}^{\perp d}>0$. Both signs have been confirmed by a flavor
analysis based on pions produced in a SIDIS experiment 
by the {\sc Hermes} \cite{hermes} and {\sc Compass} \cite{compassproton}
collaborations
and are consistent with a vanishing isoscalar Sivers function observed
by {\sc Compass} \cite{compass}.
\section{Inclusive Single-Spin Asymmetries}
Recent inclusive Deep Inelastic Scattering (DIS) experiments on a transversely polarized target in Hall A at Jefferson Lab showed for the first time a (small) Single-Spin Asymmetry (SSA) for the scattered electron. 
As such an asymmetry has to vanish in single photon exchange, these measurements potentially reveal important
information about quark correlations in the nucleon.

In a DIS process, the transverse position of the
scattered electron should be very close to that of the struck quark. One may thus estimate the effect from the initial and final state interactions of the electron by correlating the
leading twist quark density with the electromagnetic field
strength tensor at the same transverse position. 
This observation  motivates to consider \cite{metz}
\be
-M\epsilon_T^{ij}S_T^j  F_{FT}^q \equiv 
\int \frac{d\xi^-d\zeta^-}{2(2\pi)^2}e^{ixP^+\xi^-}
\langle P,S|\bar{\psi}^q(0)\gamma^+eF_{QED}^{+i}(\zeta^-)\psi^q(\xi^-)|P,S\rangle,
\label{eq:FFT}
\ee
where $e>0$ is the elementary electric charge. $x$ represents the quark
momentum, which is diagonal in this
'soft photon pole' matrix element. If the electromagnetic field strength tensor $F_{QED}^{+i}$ is replaced by its QCD counterpart $G^{+i} $  then Eq. (\ref{eq:FFT}) represents the Qiu-Sterman
matrix element \cite{QS} for the single-spin asymmetry
in Semi-Inclusive DIS (SIDIS). We will make use of this analogy several times.

Although Eq. (\ref{eq:FFT}) represents
the average transverse momentum acquired by the
electron due to ISI and FSI, it would yield the
average transverse momentum of the active quark due to
electromagnetic FSI if we were to multiply by
$\frac{1}{2} e_q$, where $e_u=\frac{2}{3}$ and
$e_d=-\frac{1}{3}$. The factor $\frac{1}{2}$  accounts for the fact that the $e^-$
in inclusive DIS experiences both ISI and FSI, while the quark in SIDIS experiences only FSI.

As a corollary, one finds 
the 'sum rule'  \cite{mb:quark}
\be
\frac{2}{3}F_{FT}^u-\frac{1}{3}F_{FT}^d+...=0
\label{eq:sumrule}
\ee
regardless whether the target is a proton or a neutron.
Thus similar to the case for QCD\cite{mb:glue}, the average transverse momentum due to the FSI also vanishes in the
abelian case, provided one sums over all charged constituents. Note that if one neglects strange or heavier quarks, then the sum rule implies that 
$F_{FT}^u$ and $F_{FT}^d$ must have the same
sign so that they can sum to zero after weighting with $e_q$
\be
F_{FT}^d=2F_{FT}^u.
\ee
This relation is not satisfied by the quark-photon correlator in Ref. [\refcite{metz}], where
for the proton $F_{FT}^d$ and $F_{FT}^u$ have opposite signs. Below we shall explain in detail how to correct that model.

\begin{figure}
\unitlength1.cm
\begin{picture}(10,3)(2.9,14)
\includegraphics{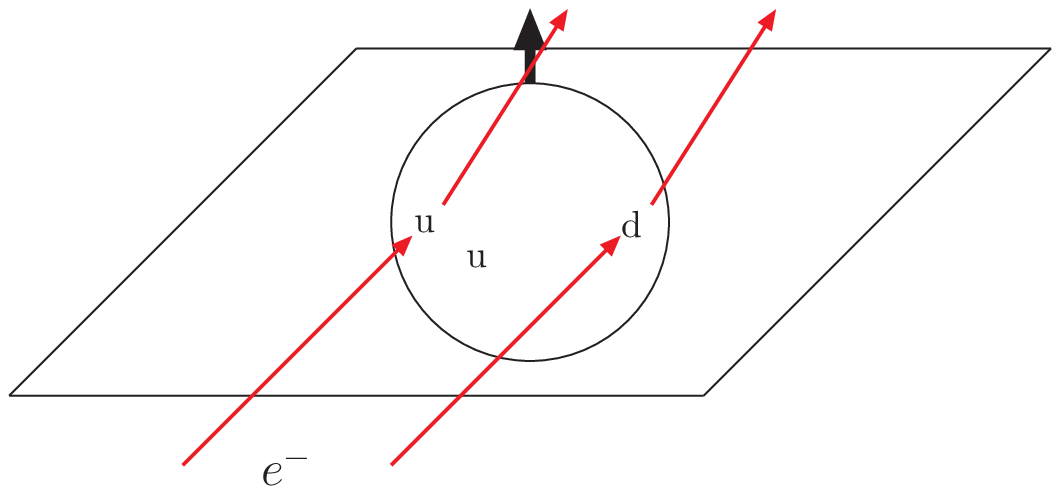}
\end{picture}
\begin{picture}(10,0)(-2.6,13.55)
\includegraphics{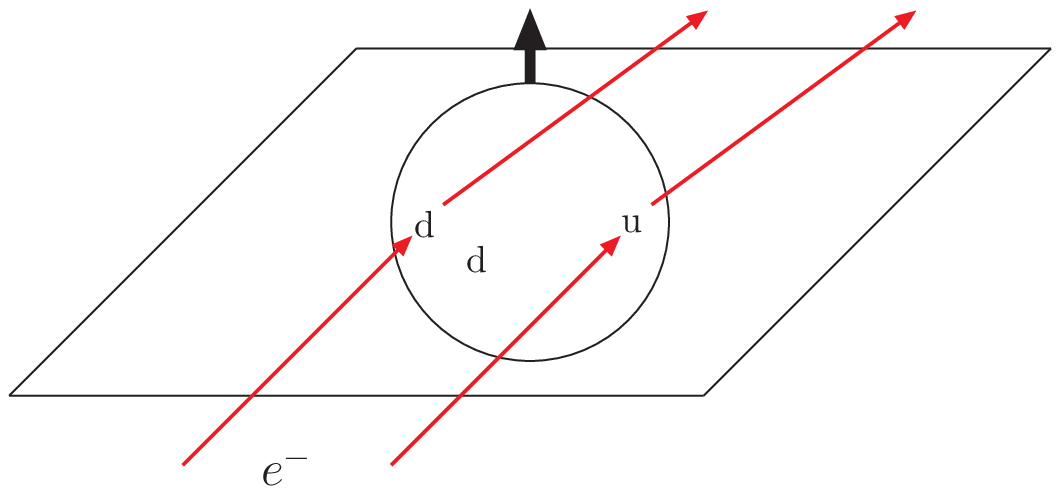}
\end{picture}
\caption{Illustration of the sign of the inclusive SSA for a proton (left) and a neutron (right) both with spin 'up'. Depending on the charge of the spectator flavor, the $e^-$ is either attracted ($u$) or repelled ($d$) by the spectators.
}
\label{fig:issa}
\end{figure}

Intuitively, the coherent contribution for scattering from $u$ {\sl vs.} $d$ quarks can be understood
from a mechanism similar to the 'lensing mechanism' proposed in Ref. [\refcite{mbSSA}].
For a transversely polarized nucleon the virtual hard photon sees $u$ quarks shifted
towards one side of the nucleon and $d$ quarks to the other. When
the $e^-$ knocks out a $u$ quark (Fig.\ref{fig:issa}), it is repelled by the negatively charged $d$ quarks
on the other side of the nucleon. As explained above the average transverse momentum from interactions with spectator $u$ quarks is zero.
On the other hand, when the $e^-$ knocks out a $d$ quark it is attracted by the
positively charged $u$ quarks. However, since the $u$ and $d$ distributions in a
transversely polarized nucleon are deformed in opposite directions, the
net force from the spectators on the $e^-$ is in both cases (knocking out
$u$ or $d$ quarks) in the same direction, i.e. there should not be a cancellation
between $u$ and $d$ quarks.

Ref. [\refcite{metz}] estimates the $\bar{q}\gamma q $ correlator in Eq. (\ref{eq:FFT}) by rescaling the $\bar{q}gq $ correlator. The latter is taken from phenomenological fits in SIDIS \cite{prokudin}. The main steps in this rescaling are multiplication by a factor $\frac{\alpha}{\alpha_s}$, multiplication by the electric charge of the spectators, and dividing by the effective color charge of the spectators.

The main reason for the violation of Eq. (\ref{eq:sumrule}) in Ref. [\refcite{metz}] is that
(by symmetry) the average transverse momentum resulting from FSI with spectators carrying the same flavor as the active quark should be zero \cite{mb:quark}, i.e.
interactions with quarks from the same flavor should be omitted when estimating the
matrix element in Eq. (\ref{eq:FFT}). This does not affect the correlation functions for the minority falvor ($d$ in proton and $u$ in neutron). However, it does affect both the QED 
ISI \& FSI as well as the QCD FSI factors when rescaling the correlation function:
The electric charge entering the QED factor should only be the electric charge of the spectator flavor ($-\frac{1}{3}$ for the $u$ in proton and $+\frac{2}{3}$ for
$d$ in the neutron). For the effective color charge in the QCD FSI, only the interaction 
with the spectator flavor should be counted, i.e. for both $u$ quarks in a proton and
$d$ quarks in a neutron that effective color charge is reduced by a factor of $2$. The resulting $\bar{q}\gamma q$
correlation functions describing inclusive SSAs are thus related to the relevant 
$\bar{q} g q$ correlation functions as (for an explanation of all factors in the rescaling see Ref. [\refcite{metz}]
\bea
F^{u/p}_{FT}(x,x)&=&\frac{\alpha_{em}}{3\pi C_F \alpha_s M}gT_F^{u/p}(x,x) \\
F^{d/p}_{FT}(x,x)&=&-\frac{2\alpha_{em}}{3\pi C_F \alpha_s M}gT_F^{d/p}(x,x)
\label{eq:newFFT1}
\\
F^{u/n}_{FT}(x,x)&=&\frac{\alpha_{em}}{3\pi C_F \alpha_s M}gT_F^{d/p}(x,x) \\
F^{d/n}_{FT}(x,x)&=&-\frac{2\alpha_{em}}{3\pi C_F \alpha_s M}gT_F^{u,p}(x,x).
\label{eq:newFFT2}
\eea
For the neutron, the resulting change of the asymmetry is small, since only
$F_{FT}^{d/n}$ has changed (increased by factor 4) compared to Ref. [\refcite{metz}].
In the asymmetry, $F_{FT}^{d/n}$ gets
multiplied by the charge squared of the down quark and thus the asymmetry increases by only about $50\%$.

For the proton the change is more significant, as our result for $F_{FT}^{u/p}$ has a sign different
from that in Ref. [\refcite{metz}], and there is
no longer an almost complete cancellation between
$u$ and $d$ contributions to $\sigma_{UT}^p$. Moreover,
since the latter gets multiplied by $e_u^2=\frac{4}{9}$, the resulting
change is quite significant. In fact, we now expect an asymmetry in the proton of the same order of magnitude as that in
the neutron. To see this we compare the cross section differences 
\be
\frac{\sigma_{UT}^p}{\sigma_{UT}^n}=
\frac{4 F_{FT}^{u/p}
+ F_{FT}^{d/p}}{4 F_{FT}^{u/n}
+ F_{FT}^{d/n}}
= 
\frac{2 T_{F}^{u/p}
- T_{F}^{d/p}}{2 T_{F}^{d/p}
- T_{F}^{u/p}}\approx -1,
\ee
where in the last step we assumed $F_{FT}^{d/p}\approx -F_{FT}^{u/p}$ consistent with a vanishing isoscalar Sivers asymmetry \cite{compass}. The predicted signs for $\sigma_{UT}$ are opposite to those in Ref. [\refcite{weiss}].

\section{Angular Momentum Decompositions}

Since the famous EMC experiments revealed that only a small fraction
of the nucleon spin is due to quark spins\cite{EMC}, 
there has been a great
interest in `solving the spin puzzle', i.e. in decomposing the
nucleon spin into contributions from quark/gluon spin and
orbital degrees of freedom.
In this effort, the Ji decomposition\cite{JiPRL}
\begin{equation}
\frac{1}{2}=\frac{1}{2}\sum_q\Delta q + \sum_q { L}_q^z+
J_g^z
\label{eq:JJi}
\end{equation}
appears to be very useful: through GPDs,
not only the quark spin contributions $\Delta q$ but also
the quark total angular momenta $J_q \equiv \frac{1}{2}\Delta q + 
{ L}_q^z$ (and by subtracting the spin piece also the
the quark orbital angular momenta $L_q^z$) entering this decomposition can be accessed experimentally. In the Ji decomposition (\ref{eq:JJi}) the quark OAM
is defined as the expectation value
\begin{equation}
{ L}_q^z= \int d^3r \langle PS| q^\dagger \left({\vec r} \times \frac{1}{i}{\vec D}
\right)^zq |PS\rangle /\langle PS|PS\rangle
\label{M012}
\end{equation}
in a nucleon state polarized in the $+\hat{z}$ direction. Here
${\vec D}={\vec \partial}-ig{\vec A}$ is the gauge-covariant
derivative.
The main advantages of this decomposition are that each term can be 
expressed as the
expectation value of a manifestly gauge invariant
local operator and that the
quark total angular momentum $J^q=\frac{1}{2}\Delta q+L^q$
can be related to GPDs\cite{JiPRL} 
and is thus accessible in deeply virtual Compton scattering and
deeply virtual meson production and can also be
calculated in lattice gauge theory. 

Jaffe and Manohar have proposed an alternative decomposition of the
nucleon spin, which does have a partonic interpretation\cite{JM}, and in which also two terms, 
$\frac{1}{2}\Delta q$ and $\Delta G$,
are experimentally accessible
\begin{equation}
\frac{1}{2}=\frac{1}{2}\sum_q\Delta q + \sum_q {\cal L}^q+
\Delta G + {\cal L}^g.
\label{eq:JJM}
\end{equation}
In this decomposition the quark OAM is defined as 
\begin{equation}
{\cal L}^q \equiv \int d^3r \langle PS|q^\dagger_+\!\left({\vec r}\times \frac{1}{i}{\vec \partial}
\right)^z \!\!q_+  |PS\rangle / \langle PS|PS\rangle
\label{M+12}
\end{equation}

\section{TMDs and OAM from Wigner Distributions}

Wigner distributions can be defined as 
off forward matrix elements of non-local
correlation functions\cite{wigner,jifeng,Metz} with $P^+=P^{+\prime}$, $P_\perp = -P_\perp^\prime = \frac{q_\perp}{2}$
\begin{eqnarray}\label{eq:wigner}
\!\!\!\!\!\!\!W^{\cal U}\!(x,\!{\vec b}_\perp,\! {\vec k}_\perp)\!
\equiv \!\!\!
\int \!\!\frac{d^2{\vec q}_\perp}{(2\pi)^2}\!\!\int \!\!\frac{d^2\xi_\perp d\xi^-\!\!\!\!}{(2\pi)^3}
e^{-i{\vec q}_\perp \!\!\cdot {\vec b}_\perp}\!
e^{i(xP^+\xi^-\!\!-{\vec k}_\perp\!\!\cdot{\vec \xi}_\perp)}
\langle P^\prime S^\prime |
\bar{q}(0)\Gamma {\cal U}_{0\xi}q(\xi)|PS\rangle .
\end{eqnarray}
Throughout this paper, we will chose ${\vec S}={\vec S}^\prime = \hat{\vec z}$. Furthermore, we will focus on the 'good' component by selecting $\Gamma=\gamma^+$.
To ensure manifest gauge invariance, a Wilson line gauge link 
${\cal U}_{0\xi}$ connecting the quark field operators at position $0$ and $\xi$ is included. The issue of choice of path
for the Wilson line will be addressed below. 

In terms  of  Wigner distributions,  TMDs and OAM can be defined 
as \cite{lorce}
\begin{eqnarray}
f(x,{\vec k}_\perp) &=& \int dx d^2{\vec b}_\perp d^2{\vec k}_\perp {\vec k}_\perp 
W^{\cal U}(x,{\vec b}_\perp,{\vec k}_\perp)\\
L_{\cal U}&=& \int dx d^2{\vec b}_\perp d^2{\vec k}_\perp \left({\vec b}_\perp \times {\vec k}_\perp \right)^z
W^{\cal U}(x,{\vec b}_\perp,{\vec k}_\perp).
\nonumber
\end{eqnarray}
No issues with the Heisenberg uncertainty principle arise here since only perpendicular combinations of position ${\vec b}_\perp$ and momentum ${\vec k}_\perp$ are
needed simultaneously in order to evaluate the integral for
$L_{\cal U}$.

A straight line connecting $0$ and $\xi$ for the Wilson line in ${\cal U}_{0\xi}$ results in
\cite{jifeng}
\begin{eqnarray}
L^q_{straight}
&=&
L^q_{Ji}.
\label{eq:LJi}
\end{eqnarray}
However, depending on the context, other choices for the path in the Wilson link ${\cal U}$ should be made. Indeed for TMDs probed in SIDIS the path should be taken to be a straight line to $x^-=\infty$
along (or, for regularization purposes, very close to) the light-cone. This particular choice ensures proper inclusion of the FSI experienced by the struck quark as it leaves the nucleon
along a nearly light-like trajectory in the Bjorken limit. However, a Wilson line to
$\xi^-=\infty$, for fixed ${\vec \xi}_\perp$ is not yet sufficient to render Wigner distributions
manifestly gauge invariant, but a link at $\xi^-=\infty$ must be included to ensure manifest
gauge invariance. While the latter may be unimportant in some gauges, it is crucial in
light-cone gauge for the description of TMDs relevant for SIDIS \cite{jifengTMD}. 

\begin{figure}
\unitlength1.cm
\begin{picture}(10,2.2)(1.2,19)
\includegraphics{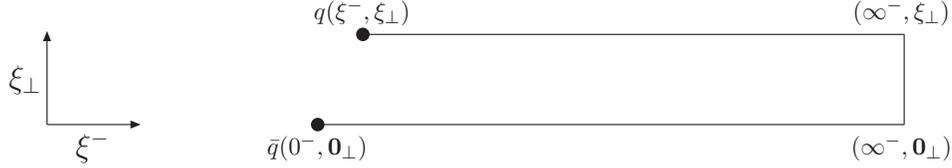}
\end{picture}
\caption{Illustration of the path for the Wilson line gauge link ${\cal U}^{+LC}_{0\xi}$ entering  $W^{+LC}$ 
(\ref{eq:wigner}).
}
\label{fig:staple}
\end{figure}

Let ${\cal U}^{+LC}_{0\xi}$ be the Wilson path ordered exponential obtained by first taking
a Wilson line from $(0^-,{\vec 0}_\perp)$ to $(\infty,{\vec 0}_\perp)$, 
then to $(\infty,{\vec \xi}_\perp)$, and then to $(\xi^-,{\vec \xi}_\perp)$, with each segment being a straight line (Fig. \ref{fig:staple}) \cite{hatta}. 
The shape of the segment at $\infty$ is irrelevant as the gauge field is pure gauge there, but it is still necessary to include a connection at $\infty$ and for
simplicity we pick a straight line. Likewise, with a similar 'staple' to $-\infty$ we define the Wilson path ordered exponential ${\cal U}^{-LC}_{0\xi}$, and using those light-like 
gauge links\footnote{Subtleties in regularizing/renormalizing such objects are addressed in Ref. \refcite{Collins:what}.}, we define
\begin{eqnarray}\label{eq:wignerpm}
\!\!\!W^{\pm LC}\!(x,\!{\vec b}_\perp,\! {\vec k}_\perp\!)\!
\equiv\!\!\! \int \!\!\!\frac{d^2{\vec q}_\perp}{(2\pi)^2}\!\!\!\int\!\!\! \frac{d^2\xi_\perp d\xi^-\!\!\!\!}{(2\pi)^3}
e^{-i{\vec q}_\perp\!\! \cdot {\vec b}_\perp}\!
e^{i(xP^+\xi^-\!\!-{\vec k}_\perp\!\!\cdot{\vec \xi}_\perp)}
\!\langle P^\prime\! S^\prime |
\bar{q}(0)\Gamma {\cal U}^{\pm LC}_{0\xi}\!\!q(\xi)|P\!S\rangle .
\end{eqnarray}
This definition for $W^{+LC}$ the same as that in Ref. \refcite{hatta} and similar to that of $W_{LC}$ in Ref. [\refcite{jifeng}] (the link segment at $\xi^-=\infty$ was not included in the definition of $W_{LC}$). 

In light-cone gauge $A^+=0$, only the segment at $\xi^-=\pm \infty$ contributes and 
the OAM looks similar to the local manifestly gaguge invariant expression, except
\be
{\vec r}\times {\vec A}({\vec r}) \longrightarrow {\vec r}\times {\vec A}(r^-=\pm \infty, {\bf r}_\perp).
\ee
From  PT invariance one finds that ${\cal L}_+^q={\cal L}_-^q$ \cite{hatta}.
In the Bashinsky-Jaffe definition of OAM ${\cal L}_{BJ}^q$\cite{BJ}\!, the vector potential in the gauge covariant derivative is replaced by
\be
\frac{\int_{-\infty}^\infty dx^- A_\perp (r^-,{\bf r}_\perp)}{\int_{-\infty}^\infty dx^-} = \frac{1}{2}\left[ { A}_\perp (r^-= \infty, {\bf r}_\perp)+
{ A}_\perp (r^-= \infty, {\bf r}_\perp)\right],
\ee
and is thus equivalent to the light-cone-staple definition
\be
{\cal L}_{BJ}^q = \frac{1}{2}\left({\cal L}_+^q+{\cal L}_-^q\right) ={\cal L}_+^q={\cal L}_-^q.
\ee
Imposing $A^+=0$ does not completely fix the gauge as one can still make $r^-$-independent gauge transformations.
If one fixes this residual gauge invariant by imposing anti-symmetric boundary conditions
$A_\perp(r^-=-\infty,{\bf r}_\perp)=-A_\perp(r^-=-\infty,{\bf r}_\perp)$ the vector potential
at $r^-=\pm \infty$ cancels out in ${\cal L}_+^q+{\cal L}_-^q$ and therefore, with the
understanding of anti-symmetric boundary conditions at $r^-=\pm \infty$ the Jaffe-Manohar OAM
becomes also identical to ${\cal L}^q_\pm$. 

This last observation is crucial for understanding the difference 
between the Ji vs. Jaffe-Manohar OAM, which in light-cone gauge\footnote{As $L^q$ involves a manifestly gauge invariant local operator, it can be evaluated in any gauge.}
involves only the replacement ${ A}_\perp^i({\vec r}) \longrightarrow {A}_\perp^i(r^-=\pm \infty, {\bf r}_\perp)$.
Using
\begin{eqnarray}
{A}^i_\perp (r^-\!\!=\infty,{\bf r}_\perp)-{ A}^i_\perp (r^-,{\bf r}_\perp)
=\!\!\int_{r^-}^\infty\!\!\!\!\! dz^-
\partial_- {A}^i_\perp (z^-,{\vec r}_\perp)
= \!\!\int_{r^-}^\infty\!\!\!\!\! dz^- G^{+i}(z^-,{\vec r}_\perp)
\label{eq:kp}
\end{eqnarray}
where $G^{+\perp}=\partial_-A^\perp$ is the gluon field strength tensor in $A^+=0$ gauge. Note that 
\begin{equation}
-\sqrt{2}gG^{+y}\equiv -gG^{0y}-gG^{zy} = g\left(E^y-B^x
\right)
=g\left({\vec E}+{\vec v}\times {\vec B}\right)^y
\end{equation}
yields the $\hat{y}$ component of the color Lorentz force acting on a particle that moves with the velocity of light in the $-\hat{z}$ direction (${\vec v}=(0,0,-1)$) --- which is the direction of the 
momentum transfer in DIS \cite{QS,mb:force}. Thus the difference between the Jaffe-Manohar and Ji\footnote{Here we replaced $\gamma^0\rightarrow\gamma^+$ in $L^q$ as discussed in Ref. [\refcite{BC}].} OAMs
\be
{\cal L}^q-L^q = -g \!\!\int \!\!d^3x\!\left\langle P\!,\!S\right|\!
\bar{q}({\vec x})\!\gamma^+\!\!
\left[{ {\vec x}\! \times\! \!
\int_{x^-}^\infty \!\!\!\!\!dr^- F^{+\perp}(r^-,{\bf x}_\perp)
}\right]^z\!\!\!\!
q({\vec x}) \!\left| P\!,\!S\right\rangle/ \langle PS|PS\rangle
\label{eq:torque}
\ee
has the semiclassical interpretation of the change in OAM due to the torque from the FSI as the quark leaves the target:\cite{mb:torque}
while $L^q$ represents the local and manifestly gauge invariant OAM of the
quark {\it before} it has been struck by the $\gamma^*$, ${\cal L}^q$ represents 
the gauge invariant OAM {\it after} it has left the nucleon and moved to $r^-=\infty$.

\begin{figure}
\unitlength1.cm
\begin{picture}(10,7.5)(0,13.2)
\includegraphics{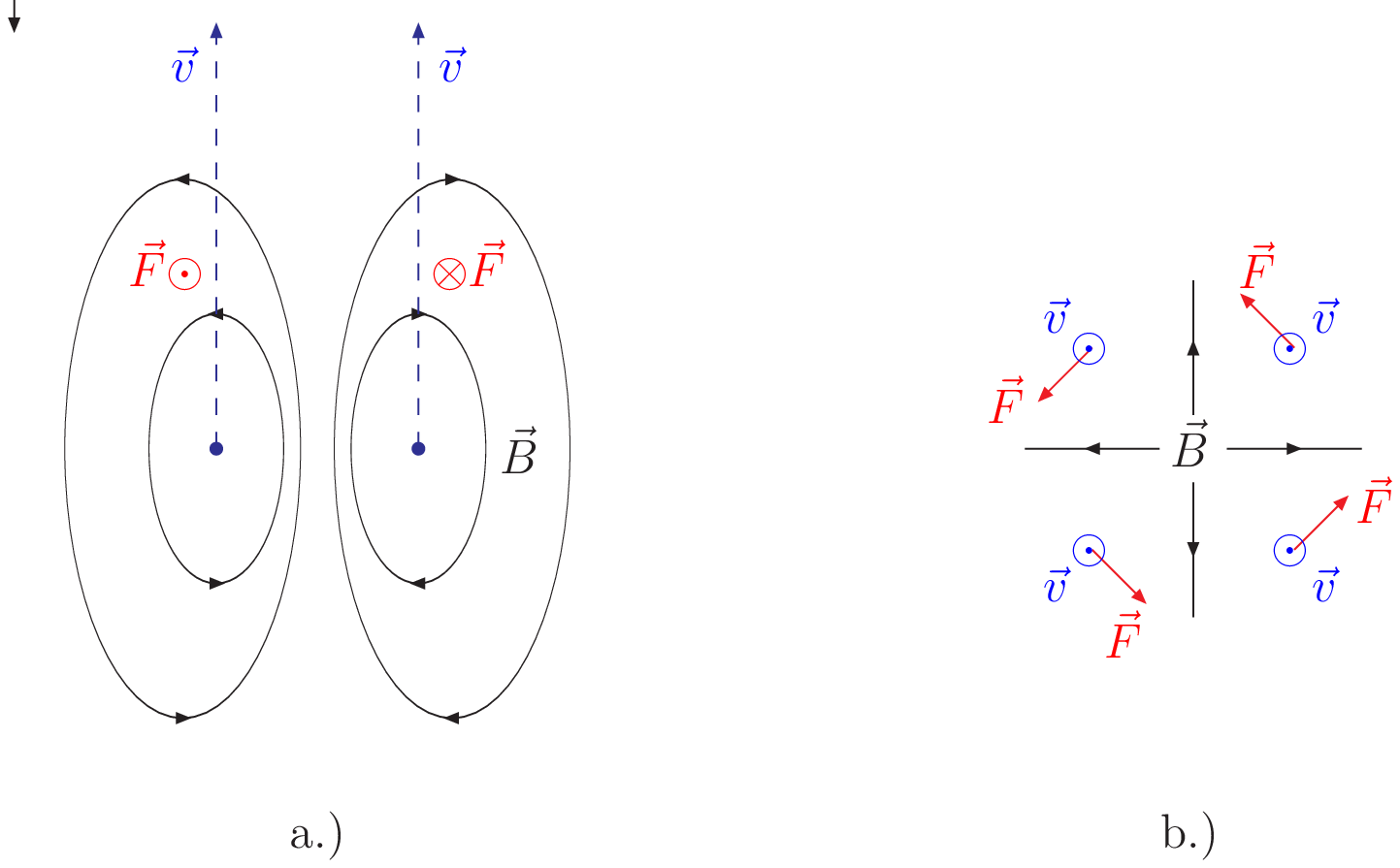}
\end{picture}
\caption{Illustration of the torque acting on the struck quark in the $-\hat{z}$ direction
through a color-magnetic dipole field caused by the spectators. a.) side view; b.) top view.
In this example the $\hat{z}$ component of the torque is negative as the quark leaves the
nucleon.
}
\label{fig:dipole}
\end{figure}


It is easy to see that a torque as appearing in (\ref{eq:torque}) may exist by considering the example of a quark moving through a (color-) magnetic dipole field caused by the spectators. Because of the overall color-neutrality, this is similar to a positively charged particle moving through the magnetic field caused by negative
spectators in QED. For spectator spins/OAMs that are oriented in the $+\hat{z}$ axis one would thus expect a dipole field as shown in Fig. \ref{fig:dipole}.
All quarks ejected in the $-\hat{z}$ direction pass through the region of outward pointing radial
magnetic field component, but only those originating in the bottom portion also move through
regions of inward pointing radial component, i.e. for quarks ejected in the $-\hat{z}$ direction
the regions of outward pointing radial component dominate.
One would thus expect more torque in the
$-\hat{z}$ direction than in $+\hat{z}$ direction. 



The observation that ${\cal L}^q={\cal L}^q_{+LC}$ is also crucial for lattice calculations of
${\cal L}^q$: In Ref. \refcite{engel}, forward matrix elements of space-like staples in fast-moving hadrons have been used for lattice calculations of TMDs including FSI. By taking nonforward matrix elements of the same operators
at small $\perp$ momentum transfer would enable to study ${\cal L}^q$.

\section{Summary}
We identified a flaw that appears in spectator models\cite{spectator} for SSAs when one considers the FSI on a quark 
from the majority flavor and corrected a phenomenological model for inclusive SSAs.

The OAM appearing in the Jaffe-Manohar formalism is identical to Wigner function based definitions of OAM utilizing light-cone staples for Wilson-line gauge links. We have used this result to
understand the difference between the Jaffe-Manohar definition of OAM and Ji's
local manifestly gauge invariant definition of OAM can be related to the torque that acts on a quark in longitudinally polarized DIS. In other words., while one definition (Ji) yields the net OAM quarks {\it before} absorbing the virtual photon, the (light-cone staple) Wigner distribution based definition (JM) yields the net OAM after the quark has escaped to infinity.
We thus now understand the physics through which  these two definitions are
related to one another. 

This is very similar to the situation in the context of TMDs where the difference between the average quark transverse momentum after it has left the target
(from Sivers function) and before it has left the target (where it is zero),
can be related to the difference of TMDs defined with a light-cone staple shaped
Wilson line gauge link versus one defined with a straight-line gauge link.

Unfortunately, no experiment has been identified to measure the OAM of quarks after they have been ejected in DIS. Nevertheless, we believe that the above interpretation
will help to develop a more complete picture of the nucleon spin. 

\section*{Acknowledgments:}
I would like to thank G. Schnell for stimulating 
discussions during the early stages of this work.
This work was partially supported by the DOE (DE-FG03-95ER40965).





\end{document}